# Unique defect structure and advantageous vortex pinning properties in superconducting CaKFe$_4$As$_4$

Shigeyuki Ishida[1*], Akira Iyo[1], Hiraku Ogino[1], Hiroshi Eisaki[1], Nao Takeshita[1], Kenji Kawashima[2], Keiichi Yanagisawa[3], Yuuga Kobayashi[3], Koji Kimoto[3], Hideki Abe[3], Motoharu Imai[3], Jun-ichi Shimoyama[4], and Michael Eisterer[5]

[1]Electronic and Photonics Research Institute, National Institute of Advanced Industrial Science and Technology, Tsukuba, Ibaraki 305-8568, Japan

[2]IMRA Materials R&D Co., Ltd., Kariya, Aichi 448-0032, Japan

[3]National Institute for Materials Science, Tsukuba, Ibaraki 305-0047, Japan

[4]Department of Physics and Mathematics, Aoyama Gakuin University, Sagamihara, Kanagawa 252-5258, Japan

[5]Atominstitut, TU Wien, Stadionallee 2, 1020 Vienna, Austria

*corresponding author: s.ishida@aist.go.jp

The lossless current-carrying capacity of a superconductor is limited by its critical current density ($J_c$). A key to enhance $J_c$ towards real-life applications is engineering defect structures to optimize the pinning landscape. For iron-based superconductors (IBSs) considered as candidate materials for high-field applications, high $J_c$ values have been achieved by various techniques to introduce artificial pinning centres. Here we report extraordinary vortex pinning properties in CaKFe$_4$As$_4$ (CaK1144) arising from the inherent defect structure. Scanning transmission electron microscopy revealed the existence of nanoscale intergrowths of the CaFe$_2$As$_2$ phase, which is unique to CaK1144 formed as a line compound. The $J_c$ properties in CaK1144 are found to be distinct from other IBSs characterized by a significant anisotropy with respect to the magnetic field orientation as well as a remarkable pinning mechanism significantly enhanced with increasing temperature. We propose a comprehensive explanation of the $J_c$ properties based on the unique intergrowths acting as pinning centres.

**INTRODUCTION**

Loss-free electrical transport is a unique property of superconductors that is utilized in various superconductivity applications. The figure of merit for the current-carrying capacity of a superconductor is $J_c$, which is determined by the material's ability to trap vortices, namely, vortex pinning.[1] Consequently, $J_c$ strongly depends on the defect structure where superconductivity is locally suppressed, and the vortices have smaller energy and are therefore pinned. Thus, how to design and introduce defects is one of the key issues towards real-life applications. To date, various techniques have been developed to control defect structures, particularly through the research on high-transition-temperature (high-$T_c$) cuprate superconductor YBa$_2$Cu$_3$O$_7$ (YBCO) thin films.[2–5] For example, nanoparticles/nanorods can be incorporated by alternately depositing YBCO and a non-superconducting (non-SC) secondary phase (e.g. Y$_2$BaCuO$_5$)[6] or by adding appropriate impurities (e.g. BaZrO$_3$) to the deposition target.[7] Moreover, stacking faults and intergrowths (e.g. extra Y or CuO planes) are frequently generated near the inclusions.[8,9] Additionally, controlled artificial defects can be created by particle irradiation,[10–12] although this technique needs complex and dedicated facilities. In any case, in order to achieve suitable defect structures, the optimization of fabrication conditions such as starting chemical composition, substrate, growth temperature, growing rate, and atmosphere is indispensable, which requires tremendous efforts. Similarly, various techniques have been exploited to introduce artificial defects in iron-based superconductors (IBSs) since their discovery.[13,14] As in the case of YBCO, $J_c$ has been enhanced particularly for $AE$Fe$_2$As$_2$-based ($AE$: alkaline-earth element) superconductors, the so-called 122 materials, by particle irradiation,[15] addition of BaZrO$_3$,[16,17] fabrication of superlattices,[18] and introduction of stacking faults.[19,20] By devising the fabrication process, a significant progress has been achieved in improving $J_c$ of bulks and thin films so far, while further $J_c$ enhancement is required towards real-life applications.

Among the 122 materials, $AE_{1-x}A_x$Fe$_2$As$_2$ ($A$: alkali metal element) possesses the highest $T_c$ up to 38 K and largest upper critical fields ($H_{c2}$) over 100 T with low anisotropy ($\gamma$) ~ 1–2. These properties are advantageous for high-field applications.[21–23] In $AE_{1-x}A_x$Fe$_2$As$_2$ (e.g. Ba$_{1-x}$K$_x$Fe$_2$As$_2$ (BaK122), Figure 1a), superconductivity is induced by substituting $AE$ with $A$ (hole doping), where $AE$ and $A$ randomly occupy the same crystallographic site in an arbitrary ratio $x$. Therefore, the superconducting properties, particularly $J_c$, of $AE_{1-x}A_x$Fe$_2$As$_2$ significantly depend on $x$.[24] Note that the significant doping dependence of $J_c$ is common to other 122 materials with different dopant elements.[25–27] As a result, a fine adjustment of $x$ is required to achieve better properties of bulks and thin films. In this study, we focus on the recently discovered 1144 materials,[28,29] $AEA$Fe$_4$As$_4$, which possess $T_c$ and $H_{c2}$ comparable to 122 materials. In the 1144 structure (Figure 1b), $AE$ and $A$ do not occupy the same site owing to the large difference in the ionic radii (e.g. 1.21 Å and 1.51 Å for Ca$^{2+}$ and K$^+$, respectively); hence, $AE$ and $A$ layers stack alternately along the $c$ axis. Therefore, the 1144 material is a line compound where the Fe valence state is fixed at 2.25+. This characteristic is advantageous for applications because fluctuations in chemical composition is, in principle, not allowed. Meanwhile, for such an ordered-layered structure, 122 phases ($AE$122 and $A$122) intergrow in the CaK1144 matrix if excess of $AE$ or $A$ prevails during the synthesis process. Since $AE$122 are non-SC parent materials and $A$122 are superconductors with low $T_c$ < 4 K (practically non-SC), such intergrowths possibly act as vortex-pinning centres. In fact, recent studies on vortex pinning properties of CaK1144 reported unusually high $J_c$[30] as well as vortex dynamics distinct from 122 materials,[31] while the relevant pinning mechanisms remain unsolved. This motivated us to explore the microstructure and the vortex pinning mechanisms in CaK1144. Here we demonstrate the unique defect structure in CaK1144, which provides comprehensive explanations of the sublime vortex pinning properties.

## RESULTS

### Microstructure of CaKFe$_4$As$_4$ single crystal

The crystal structure of the CaK1144 matrix and the unique defect structure can be directly observed by high-resolution scanning transmission electron microscopy (STEM) experiments. Figure 1c shows a low-magnification annular-dark-field (ADF)-STEM image taken along the [100] axis. Overall, the STEM image shows a uniform contrast, indicative of good homogeneity of the matrix region. Notably, characteristic bright stripes in the horizontal direction with typical lengths of ~1 μm can be identified. These structures are regarded as planar defects along the *ab* plane, while no other defects are detected. Figure 1d shows the ADF-STEM image around one of the bright stripes. The upper right panel shows the magnified view of the CaK1144 matrix. The brightest zig-zag arrangements of dumbbells indicated by green arrows are assigned to FeAs layers. The Fe-Fe interplane distance across the two kinds of relatively dark layers (the brighter and the darker ones indicated by blue and orange arrows, respectively) was determined to be 6.1 Å and 7.0 Å, respectively. These values are in good agreement with the reported ones (6.12 Å and 6.70 Å, see Figure 1b), indicating that the brighter and darker layers correspond to Ca and K layers, respectively. Thus, we confirmed that the alternating stacking of Ca and K layers is indeed realized in the matrix.

Next, we focus on the bright stripe magnified in the right lower panel in Figure 1d. It reveals that the alternation of Ca and K layers is violated, while the local FeAs-layer structure is maintained. There are nine FeAs-to-FeAs units with a total thickness of about 55 Å, and each Fe-Fe interplane distance is found to be 6.1 Å, which is identical to that across the Ca layer. The chemical composition analysis shows that Ca is rich around the defect without significant changes for Fe and As (see Supplementary information). Based on the results, we conclude that the defect is a Ca122 intergrowth with dimensions of ~5.5 nm (~5 unit cells) along the *c* axis and ~1 μm along the *ab*-plane, which is coherently grown in the CaK1144 matrix.

Furthermore, when the microstructure of CaK1144 was carefully investigated, we found much smaller defects. In Figure 1e, there is a thin bright line indicated by a black arrow. This defect is identified to be a monolayer Ca122 intergrowth, as shown in the right panel. Typically, such thin intergrowths have dimensions of 1–2 nm in thickness (along the *c*-axis) and 50–100 nm in length (along the *ab*-planes). Thus, the existence of Ca122 intergrowths with various sizes is revealed. Such intergrowths should have significant influence on the vortex pinning properties in CaK1144.

### Critical current properties in CaKFe$_4$As$_4$

Magnetization hysteresis loops (MHLs) were investigated ($M \sim J_c$) in order to explore the vortex pinning properties in CaK1144. First, MHLs with $H$ parallel to the *c*-axis are shown (Figures 2a and 2b), which have been intensively investigated to evaluate the in-plane $J_c$ for $H // c$ ($J_c^{H//c}$) in the IBSs. Each MHL shows a peak around self-field ($H$ = 0) commonly seen for other IBSs. The $H$ dependence of $J_c^{H//c}$ calculated from the MHLs is shown in Figure 2c. Surprisingly, $J_c^{H//c}$ *increases* with increasing temperature ($T$) in the high $H$ region, which is contrary to the common knowledge about the $T$ dependence of $J_c$. For example, $J_c^{H//c}$ at $T$ = 3 K (black) and that at $T$ = 20 K (light blue) cross each other at around $\mu_0 H \sim 4$ T, resulting in a larger $J_c^{H//c}$ for $T$ = 20 K in a high $H$ region. The feature is clearly seen in the $T$ dependence of $J_c^{H//c}$ ($J_c^{H//c}$ – $T$ at 0.4, 1, 3, and 6 T) plotted in Figure 2d, showing a broad peak at around 20 K under various $H$. Under the low $H$ below 1 T, the peak in $J_c^{H//c}$ – $T$ is absent since $J_c$ at low $T$ is dominated by the strong pinning contribution (corresponding to the peak around $H = 0$) arising from sparse and large pointlike defects[32] (see the Supplementary information). To our knowledge, such a large increase in $J_c$ with increasing $T$ in wide $T$ and $H$ ranges has not been reported previously in other IBSs nor high-$T_c$ cuprates (note that there are several examples of increase in $J_c$ with increasing $T$, while they are in general accompanied by a prominent second magnetization peak in MHLs[27] in contrast to the present moderate $H$ dependence of $J_c$). The unusual $T$ dependence of $J_c$, namely, the "peak effect" in $J_c$ – $T$, highlights a remarkable enhancement of pinning with increasing $T$ even at temperatures well below $T_c$, which is unique to CaK1144. It is evident that the $T$ dependence of $J_c^{H//c}$ of CaK1144 is distinct from that of the 122 materials. In Figure 2e, the $T$ dependence of $J_c^{H//c}$ at 6 T for CaK1144 is compared with those for various 122 materials[27]; $Ba_{1-x}K_xFe_2As_2$, $Ba(Fe_{1-x}Co_x)_2As_2$, and $BaFe_2(As_{1-x}P_x)_2$ with different $x$ values. Although $J_c^{H//c}$ of CaK1144 is relatively small at low $T$, the maximum $J_c^{H//c} = 0.17$ MA/cm$^2$ at 20 K is comparable to the highest one reported for 122 materials. Such high $J_c$ demonstrates that the $T$-enhanced pinning centres trap vortices very efficiently.

Next, we show MHLs with $H$ along the $ab$ plane to evaluate $J_c$ for $H // ab$ ($J_c^{H//ab}$). Figure 3a shows the MHLs for CaK1144. The shape of the MHL is clearly different from that for $H // c$ in that it shows a dip structure around self-field, which will be discussed later. Moreover, the size of the MHL monotonically decreases with increasing $T$ in contrast to the case of $H // c$, suggesting a significant anisotropy in the vortex pinning properties with respect to the $H$ orientation. Figure 3b shows the $H$ dependence of $J_c^{H//ab}$ derived from the MHLs. Here, we applied a simplified calculation procedure following the previous work[30] (see Methods and Supplementary information). The estimated $J_c^{H//ab}$ is extremely large, 5 MA/cm$^2$ at 5 K and 3 T, which is ~40 times larger than $J_c^{H//c}$. $J_c^{H//ab}$ maintains large values at higher $T$, over 1 MA/cm$^2$ up to 6 T at 20 K and up to 3 T at 25 K. These values are comparable with the highest $J_c$ in IBS thin films[20,33] (see Figure 3e).

The unusually high $J_c^{H//ab}$ in CaK1144 can be confirmed by comparing with the results of BaK122 obtained by the same procedure. Figure 3c shows the MHLs for BaK122 ($x = 0.4$). In contrast to the case of CaK1144, the MHLs show a peak around self-field, similar to that for $H // c$. Figure 3d shows the $H$ dependence of $J_c^{H//ab}$ for BaK122. Notably, $J_c^{H//ab}$ of BaK122 is much smaller than that estimated for CaK1144. For example, $J_c^{H//ab} = 0.3$ MA/cm$^2$ at 5 K and 3 T is smaller by one order of magnitude and 0.01 MA/cm$^2$ at 25 K is smaller by two orders. Such a large difference supports the high $J_c$ arising from a unique pinning mechanism in CaK1144.

Figures 3e and 3f show the $T$ dependences of $J_c^{H//ab}$ (filled circles), $J_c^{H//c}$ (open circles), and the $J_c$ anisotropy defined as $J_c^{H//ab}/J_c^{H//c}$ (stars) for CaK1144 (red symbols) and BaK122 (blue symbols), respectively. $J_c^{H//ab}$ and $J_c^{H//c}$ of Co-Ba122[20] and NdFeAs(O,F)[33] (Nd1111) thin films are plotted for comparison. In the case of CaK1144, the pinning is significantly anisotropic with respect to the $H$ orientation, as demonstrated by the distinct magnitude as well as the $T$ dependence of $J_c$ for $H // ab$ and $c$. The anisotropy tends to increase with decreasing $T$, taking a maximum value of ~40 around 5–7 K. At higher $T > 10$ K, where $J_c^{H//c}$ increases, the anisotropy decreases to ~10 at 20 K. In contrast, for BaK122, $J_c^{H//ab}$ and $J_c^{H//c}$ almost overlap each other, i.e. $J_c$ is isotropic at $T$ below 20 K. Moreover, the anisotropy increases with $T$ in contrast to the case of CaK1144. Apparently, the $T$ dependence of $J_c$ anisotropy of BaK122 is similar to that of $H_{c2}$ anisotropy.[34,35] Such a correlation between the $J_c$ anisotropy and the $H_{c2}$ anisotropy can be qualitatively understood in terms of the anisotropy of coherence length ($\xi$) together with pinning by random point defects[3]. Thus, in the case of BaK122, common pinning centres likely dominate $J_c$ both for $H \mathbin{/\mkern-3mu/} ab$ and $c$.

**DISCUSSION**

The extraordinary vortex pinning properties in CaK1144 are summarized as follows; (i) $J_c^{H//c} - T$ shows an unexpected peak effect and (ii) $J_c^{H//ab}$ is unusually large. Regarding $H \mathbin{/\mkern-3mu/} c$, the $H$ and $T$ dependence of $J_c^{H//c}$ of CaK1144 is visualized in the form of a contour plot in Figure 4a. Several characteristic $T$ and $H$ corresponding to the two types of peak effect observed in $J_c^{H//c} - T$ and MHLs are also marked. For comparison, the corresponding data for BaK122 ($x = 0.3$) which possesses the highest $J_c$ among the 122 materials[27] are shown in Figure 4b. The colour distribution for CaK1144 is characterized by the hot-colour region in the intermediate $T$ range. It is found that the peak in $J_c^{H//c} - T$ ($T_p$) is almost $H$-independent, suggestive of a unique origin of the enhanced pinning with increasing $T$. At high $T$ region (approximately above $T_p$), the peak in MHLs ($H_p$) appears in the observable $H$ range (< 7 T) similarly to BaK122, which is in general associated with the order-disorder transition of the vortex lattice. It is evident that $T_p$ and $H_p$ are well-separated in the $H - T$ phase diagram, suggestive of the different mechanisms underlying the two types of peak effect.

Now we return to the defect structure in CaK1144 to understand the anomalous $J_c$ properties. The Ca122 intergrowths observed by the STEM are schematized in Figures 1f and 1g. The colour gradation indicates the difference in $T_c$ between the matrix and the defects. The intergrowths are considered to be categorized into two types; (i) intergrowths which are thick (5–10 nm) along $c$ axis and large (~ 1 μm) along the $ab$ plane (Figure 1f), and (ii) thin (1–2 nm) and small (~100 nm) ones (Figure 1g). For the former case, the thickness is typically ~ 5 nm, as represented by Figure 1d, which is much larger than the $c$-axis coherence length ($\xi_c \sim 1$ nm)[36] of CaK1144. Such intergrowths are regarded as non-SC planar defects because the inner part of the intergrowths is considered to be undoped Ca122. In general, these defects act as efficient pinning centres for $H \mathbin{/\mkern-3mu/} ab$ while they do not contribute to pinning for $H \mathbin{/\mkern-3mu/} c$. On the other hand, for the latter case, when the thickness is ~1–2 nm, i.e. 1-2 Ca layers are inserted (Figure 1e), holes can be supplied to the inner FeAs layers from the K layers, hence such intergrowths are considered to be SC defects. It is expected that $T_c$ of the SC defects ($T_c^{defect}$) is lower than that of the CaK1144 matrix due to the depleted carrier density as in the case of underdoped BaK122. Then, $T_c^{defect}$ is determined by the number of Ca layers in the defect, hence it likely takes discrete values. In addition, these defects terminate in a short range (< ~100 nm); hence, $T_c$ abruptly changes along the $ab$ plane around their ends. Therefore, they act as effective pinning centres not only for $H \mathbin{/\mkern-3mu/} ab$ but also for $H \mathbin{/\mkern-3mu/} c$.

Among those two types of defects, the former one is considered to give rise to the unusually large $J_c^{H//ab}$ as well as $J_c$ anisotropy as in the case of artificial superlattices in thin films. In addition, such defects can account for the dip feature in the MHLs, which has been reported for irradiated IBSs where $J_c$ is significantly enhanced. For the dip feature, two explanations have been considered: a highly inhomogeneous field distribution[37] and the anisotropy of $J_c$.[38] Both are compatible with the properties of CaK1144. The self-field is in general inhomogeneous with strongly curved flux lines, hence the local field can not be parallel to the intergrowths in the entire sample, and thus pinning by the intergrowths is less effective at low fields. Meanwhile, the intergrowths can cause the large $J_c$ anisotropy ($J_c^{H//ab} \gg J_c^{H//c}$ as well as the inter-/intra-plane $J_c$ anisotropy).

On the other hand, the latter type is considered to play a key role in the unusual $T$ dependence of $J_c^{H//c}$. The strength of pinning around the ends of the intergrowths is determined by the difference in condensation energy between the matrix and defects ($\Delta E_c = E_c^{1144} - E_c^{defect}$). The condensation energy ($E_c = H_c^2/8\pi$ where $H_c$ is thermodynamic critical field), which is the difference of the ground state energies between the normal state and the SC state, depends on $T$. Because the thin intergrowths are superconducting at low $T$ ($E_c^{defect} > 0$), $\Delta E_c$ is likely small hence the pinning is weak. When the intergrowths turn into the normal state ($E_c^{defect} = 0$) with increasing $T$, the pinning becomes stronger owing to the larger energy gain. Thus, the thin intergrowths, i.e. the SC defects, are regarded as $T$-enhanced pinning centres, which possibly give rise to the increase in $J_c^{H//c}$ with increasing $T$.

To our knowledge, the idea of SC defects has been well-known, while the $T$ dependence of $J_c$ in the presence of SC defects has not been sufficiently investigated. Here, we calculate the pinning force density $f_p$ using a simple model; $f_p \sim \Delta E_c \xi$ where $\Delta E_c = E_c^{1144} - E_c^{defect}$ is the difference in $E_c$ between CaK1144 matrix and SC defect ($\Delta E_c > 0$ considering $T_c^{defect} < T_c^{1144}$), and $\xi$ is the coherence length. Here, the $T$ dependences of $E_c$ and $\xi$ are modelled by $E_c \sim H_c^2 \sim [1 - (T/T_c)^2]^2$ and $\xi \sim [(1 + (T/T_c)^2)/(1 - (T/T_c)^2)]^{1/2}$, respectively. First, in the case of non-SC defects where $T_c^{defect} = 0$ and $E_c^{defect} = 0$ (i.e. $\Delta E_c = E_c^{1144}$), $f_p$ ($\sim E_c^{1144}\xi$) monotonically decreases with increasing $T$ (Figure 5a). Next, an example result for SC defects is shown in Figure 5b. Here, $T_c^{defect} = 25$ K (corresponding to underdoped BaK122 with $x \sim 0.25$) and $E_c^{defect}(0)/E_c^{1144}(0) = 0.6$ were chosen (for the results using other parameters, see Supplementary information). In this case, $\Delta E_c$ shows a weak $T$ dependence below $T_c^{defect}$ owing to the increase of $E_c^{defect}$. As a result, $f_p$ increases with increasing $T$, showing a peak around 20 K, which qualitatively agrees with the present observations (Figure 2d). Note that the peak position in $f_p - T$ depends on $T_c^{defect}$ (see Supplementary information), hence $T_c^{defect}$ is likely correlated with $T_p$ in the $H - T$ phase diagram (Figure 4). In addition, the vortex lattice softens with increasing $T$ which allows for a better accommodation of the lattice to the defect structure and hence triggers an order-disorder transition of the vortex lattice. This tendency is compatible with the appearance of second magnetization peak at higher temperatures in CaK1144. Thus, the unusual $T$ dependence of $J_c$ in CaK1144 can be qualitatively understood by considering the SC defects unique to this material. In the present case, the feature is pronounced possibly because (i) CaK1144 is essentially a clean system as indicated by the relatively low $J_c$ at low $T$ and (ii) Ca122 intergrowths take discrete $T_c^{defect}$ values determined by the number of Ca layers, resulting in a single peak in $T$ dependence of $J_c$. However, to quantify the influence of the Ca122 intergrowths on the unusual $T$ dependence of $J_c$, further experimental investigations such as determination of defect density as well as more detailed theoretical calculations are desired.

To summarize, we demonstrated a clear correlation between the microstructure and the vortex pinning properties of CaK1144. The nanoscale Ca122 intergrowths inherent to CaK1144 single crystals result in an unusual $T$ dependence of $J_c^{H//c}$ as well as extremely large $J_c^{H//ab}$, distinct from other IBSs. The advantageous vortex pinning properties will offer a new route for further improvement of $J_c$ and enhance the application potentiality of IBSs.

## METHODS

### Single crystal growth

Single crystals of CaK1144 were grown by the FeAs-flux method.[39] The FeAs precursor was prepared from Fe and As mixed at a ratio of 1 : 1 and heated at 900 °C for 10 h in an evacuated quartz tube. Ca, K, and FeAs were weighed at a ratio of 1 : 1.1 : 10 and placed in a zirconia crucible, then sealed in a Ta container using an arc-welding chamber. The Ta container was sealed in an evacuated quartz tube to protect Ta from oxidation. The container was heated during 5 h to 650 °C and held there for 5 h. It was then heated to 1180 °C within 5 h and held there for another 5 h. Then, it was cooled over 5 h to 1050 °C, followed by slow cooling to 930 °C for 80 h. For the single crystals used in this study, X-ray diffraction (XRD) patterns were measured at room temperature using a diffractometer with Cu K$\alpha$ radiation (Rigaku, Ultima IV) to check 00$l$ peaks (see Supplementary information). No trace of Ca122 and K122 was observed within the resolution of XRD.

### Scanning transmission electron microscopy

The microstructure of a CaK1144 single crystal was investigated using an aberration-corrected scanning transmission electron microscope (FEI, Titan cubed) at an acceleration voltage of 300 kV. The sample was prepared using a focused ion beam (Hitachi, FB-2000). The chemical composition was investigated by electron energy loss spectroscopy (EELS, Gatan, GIF Quantum ERS) and energy dispersive X-ray spectroscopy (EDS, Oxford Instruments, X-Max$^N$ 100TLE).

### In-plane resistivity measurements

The in-plane resistivity $\rho_{ab}(T)$ (shown in the Supplementary information) was measured by a standard four-probe method using a physical property measurement system (Quantum Design). Magnetic fields up to 9 T were applied along the $c$ axis and in the $ab$ plane to evaluate the anisotropy of upper critical fields. As shown in the Supplementary information, the residual resistivity ratio ($\rho_{ab}(300K)/\rho_{ab}(40K)$) was ~16, and no trace of magneto-structural phase transition of Ca122 phase was observed around 170 K. These properties meet the criteria for "phase-pure" single crystals in Ref. 39.

### Magnetization measurements

The samples for the magnetization measurements were cut into rectangular shapes. For CaK1144, the dimensions were $l$ = 1.57 mm (length), $w$ = 1.34 mm (width), and $d$ = 0.035 mm (thickness). For BaK122, the dimensions were $l$ = 1.59 mm, $w$ = 0.764 mm, and $d$ = 0.099 mm. The measurements were performed using a magnetic property measurement system (Quantum Design). For $H$ // $c$, $J_c^{H//c}$ was calculated using Bean's critical state model[40]; $J_c^{H//c}$ = 20$\Delta M$/$w$(1-$w$/3$l$) where $\Delta M$ is the width of the MHLs. For $H$ // $ab$, two $J_c$ components (in-plane $J_c$ ($J_c^{H//ab}$) and inter-plane $J_c$ ($J_c^c$)) contribute to $M$. Here, we used a simplified formula for the evaluation of $J_c^{H//ab}$ by taking $J_c^{H//ab} = J_c^c$, i.e. $J_c^{H//ab}$ = 20$\Delta M$/$d$(1-$d$/3$l$), following the previous study.[30] We confirmed that this simplified procedure does not alter the main conclusions in this study. For more details, see the Supplementary information where the evaluation of $J_c^{H//ab}$ and $J_c^c$ using the extended Bean's critical state model for anisotropic $J_c$[41,42] is described.

**ACKNOWLEDGEMENTS**

We thank T. Tamegai and S. Pyon for fruitful discussion on the evaluation of anisotropic $J_c$. This work was supported by the Austria-Japan Bilateral Joint Research Project hosted by the Japan Society for the Promotion of Science (JSPS) and by FWF: I2814-N36, and a Grant-in-Aid for Scientific Research (KAKENHI) (JSPS Grant Nos. JP16K17510 and JP16H06439).

**AUTHOR CONTRIBUTIONS**

The research plan was designed and coordinated by A.I., S.I., K.Kawashima, and H.E. S.I., A.I., H.O., N.T., H.A., and M.I. carried out the single crystal growth and the basic characterization of CaK1144 and BaK122. S.I. conducted the electrical transport and the magnetization measurements on the single crystals. K.Y., Y.K., and K.Kimoto performed the STEM measurements and conducted the data analysis. S.I. and M.E. carried out the numerical calculation. S.I. wrote the main body of the manuscript under the support of other coauthors, particularly by M.E., J.S., and H.E.; all authors contributed to the discussion of the results for the manuscript.

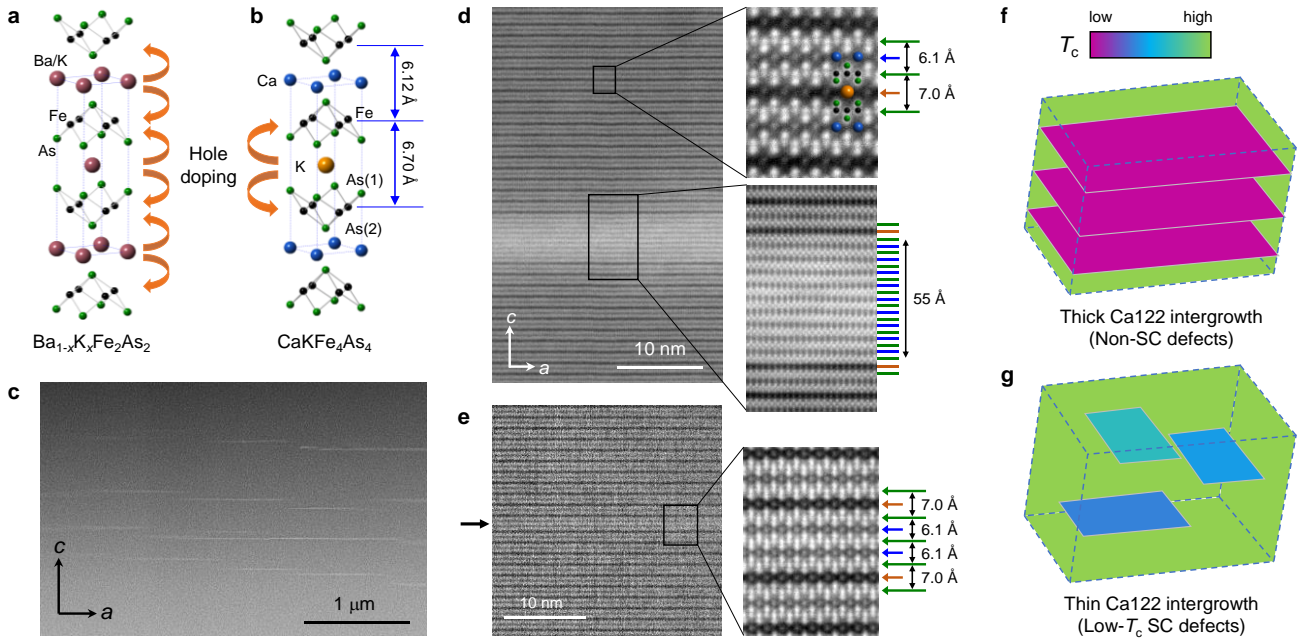

**Fig. 1** Microstructure of CaK1144 single crystal investigated by STEM. **a,b** Crystal structure of BaK122 and CaK1144. **c** STEM image with low magnification. A number of defects (bright lines) with length of ~1 μm can be identified. **d** High resolution STEM image around the defects shown in **c**. A magnified view of the CaK1144 matrix and the defect are shown in right upper and lower panels, respectively. The structural model of CaK1144 is overlapped in the upper panel in accordance with the observed STEM image. The defect with thickness of 55 Å (each Fe-Fe interplane distance is 6.1 Å) is found to be Ca122 intergrowth. We did not observe intergrowths of $KFe_2As_2$ nor FeAs (flux material) inclusions. **e** Thin defect observed by STEM and the enlarged view, demonstrating a monolayer Ca122 intergrowth. **f,g** Schematic models for thick and thin Ca122 intergrowths. Colour gradation indicates $T_c$. Thick intergrowths are regarded as non-superconducting planar defects, while thin ones are considered to be superconducting defects with a lower $T_c$ than in the CaK1144 matrix. Such intergrowths act as effective pinning centres, giving rise to the unusual $J_c$ properties in CaK1144.

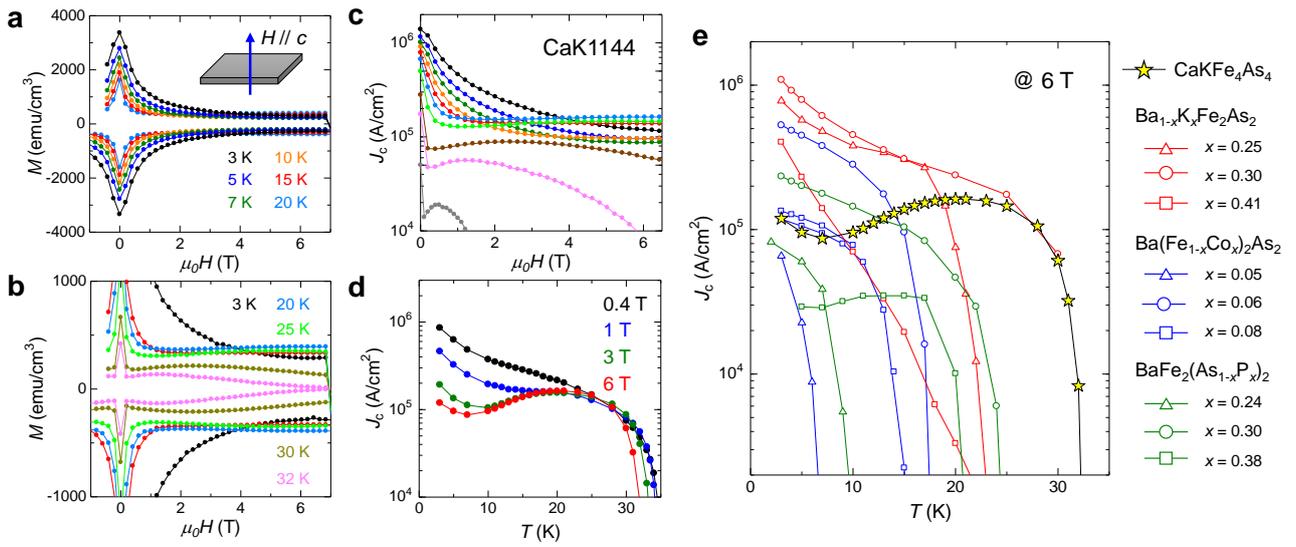

**Fig. 2** Critical current properties of CaK1144 single crystal for $H \parallel c$. **a** MHLs at $T$ = 3–20 K. **b** MHLs at $T$ = 3 K and 20–32 K. **c** Magnetic field dependence of $J_c$. **d** Temperature dependence of $J_c$ at $\mu_0 H$ = 0.4, 1, 3, and 6 T. **e** Comparison of $J_c$ with representative 122 single crystals[27], BaK122 (red), Co-Ba122 (blue), and P-Ba122 (green) with various doping concentrations $x$. In addition to CaK1144, P-Ba122 with $x$ = 0.38 shows an increase in $J_c$ with increasing $T$. In this case, however, a MHL is characterized by a sharp second magnetization peak, which is apparently different from CaK1144[27].

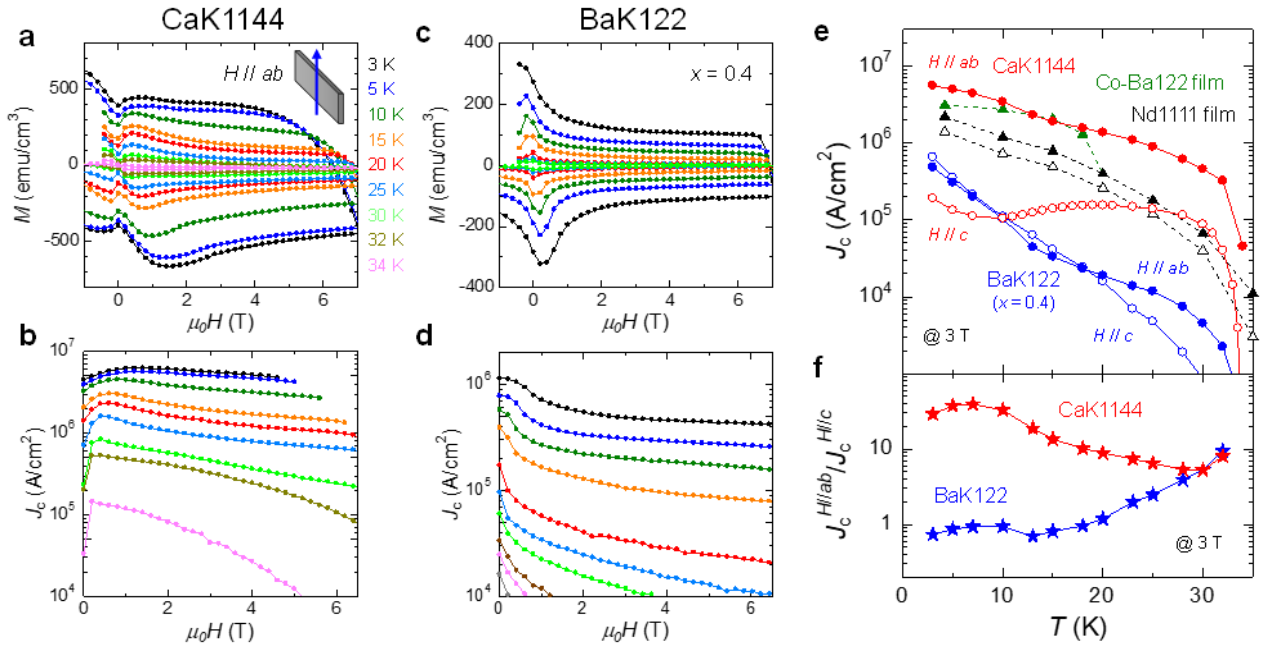

**Fig. 3** Critical current properties of CaK1144 single crystal for $H \parallel ab$ in comparison with BaK122 ($x$ = 0.4). **a,b** MHLs at $T$ = 3–34 K and magnetic field dependence of $J_c$ for CaK1144. **c,d** Same data set as in **a, b** for BaK122. **e** Temperature dependence of $J_c$ for CaK1144 (red) and BaK122 (blue) under $H \parallel ab$ (filled) and $c$ (open). $J_c$ data of Co-Ba122[20] and Nd1111[33] thin films are plotted for comparison. Note that $J_c$ of Co-Ba122 film is almost isotropic. **f** Temperature dependence of $J_c$ anisotropy defined as $J_c^{H\parallel ab}/J_c^{H\parallel c}$.

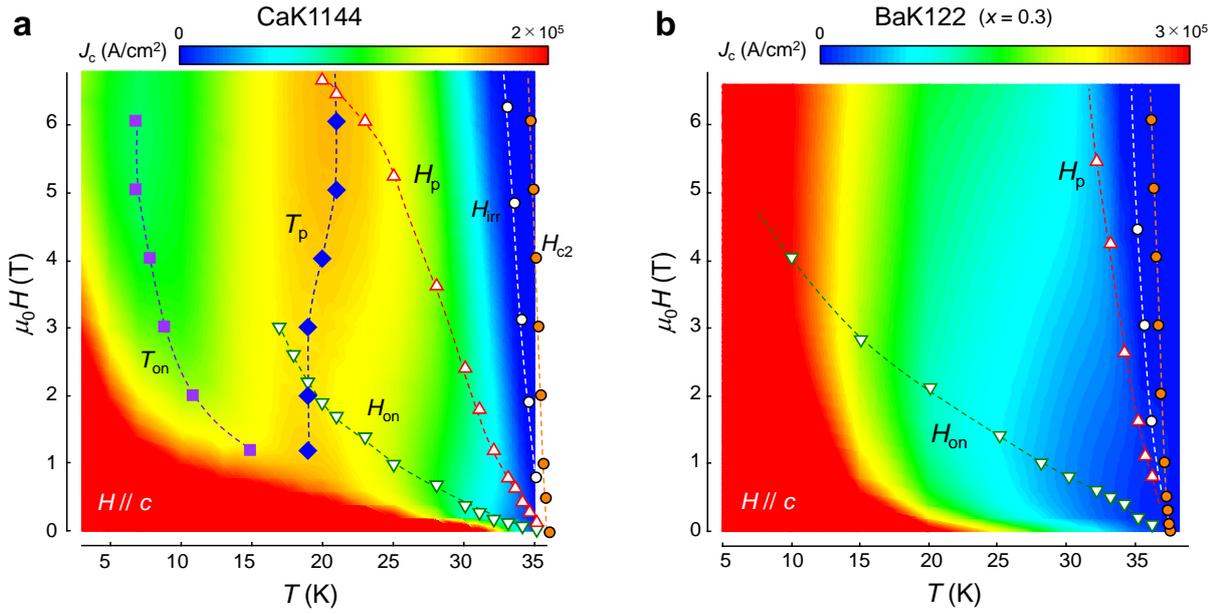

**Fig. 4** Vortex phase diagrams under $H \parallel c$ for **a** CaK1144 and **b** BaK122 ($x$ = 0.3). $H$ and $T$ dependences of $J_c$ are shown in the form of contour plots. Hot (cold) colours indicate high (low) $J_c$ region. Several characteristic $T$ and $H$ are plotted: $T_{on}$, the onset of the peak effect in $J_c$ – $T$ curves defined by the local minimum (purple squares); $T_p$, the peak position in $J_c$ – $T$ (blue diamonds); $H_{on}$, the onset of the second magnetization peak in $M$ – $H$ curves defined by the local minimum (open reversed triangles); $H_p$, the second peak position in $M$ – $H$ (open triangles); $H_{irr}$, irreversibility field defined by a criterion of $J_c <$ 100 A/cm$^2$ (open circles); and $H_{c2}$, the upper critical field along the $c$ axis obtained from the resistivity measurements (orange circles). The dashed lines are guide for the eye.

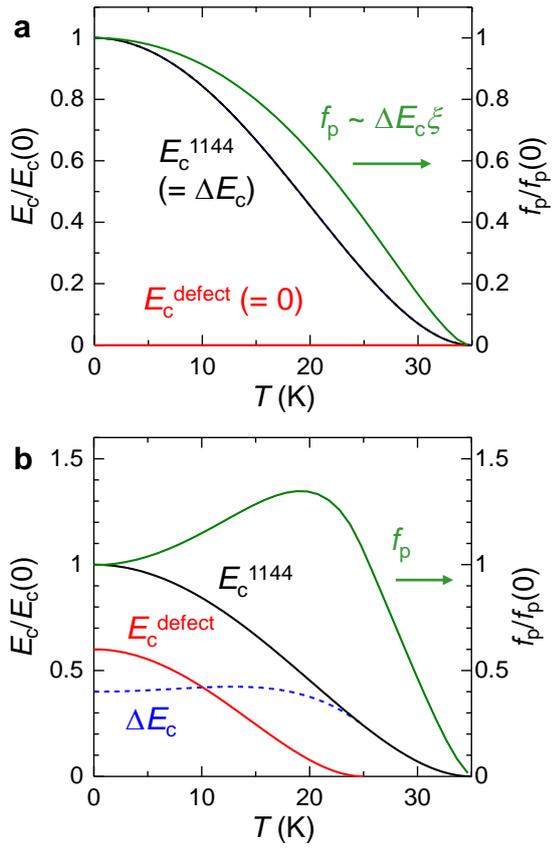

**Fig. 5** Temperature dependence of pinning force density $f_p$. **a** $f_p$ (~ $\Delta E_c \xi$, green curve) calculated for non-SC defects where $E_c^{defect} = 0$ (red curve), i.e. $\Delta E_c = E_c^{1144}$ (black curve). In this case, $f_p$ monotonically decreases with increasing $T$. **b** $f_p$ calculated for SC defects with $T_c^{defect} = 25$ K and $E_c^{defect}(0)/E_c^{1144}(0) = 0.6$. $\Delta E_c$ (blue dashed curve) shows a weak $T$ dependence below $T_c^{defect}$, resulting in an enhancement of $f_p$ with increasing $T$.

Supplementary Information for

# Unique defect structure and advantageous vortex pinning properties in superconducting CaKFe$_4$As$_4$

Shigeyuki Ishida[1*], Akira Iyo[1], Hiraku Ogino[1], Hiroshi Eisaki[1], Nao Takeshita[1], Kenji Kawashima[2], Keiichi Yanagisawa[3], Yuuga Kobayashi[3], Koji Kimoto[3], Hideki Abe[3], Motoharu Imai[3], Jun-ichi Shimoyama[4], and Michael Eisterer[5]

[1]Electronic and Photonics Research Institute, National Institute of Advanced Industrial Science and Technology (AIST), Tsukuba, Ibaraki 305-8568, Japan

[2]IMRA Materials R&D Co., Ltd., Kariya, Aichi 448-0032, Japan

[3]National Institute for Materials Science (NIMS), Tsukuba, Ibaraki 305-0047, Japan

[4]Department of Physics and Mathematics, Aoyama Gakuin University, Sagamihara, Kanagawa 252-5258, Japan

[5]Atominstitut, TU Wien, Stadionallee 2, 1020 Vienna, Austria

[*]corresponding author: s.ishida@aist.go.jp

## 1. Characterization of CaKFe$_4$As$_4$ and Ba$_{0.6}$K$_{0.4}$Fe$_2$As$_2$ single crystals

In advance of the scanning transmission electron microscopy (STEM) and magnetization hysteresis loop (MHL) measurements, the quality of CaKFe$_4$As$_4$ (CaK1144) single crystals was checked by X-ray diffraction (XRD), magnetization ($M$) and in-plane resistivity ($\rho_{ab}$) measurements following the procedure in Ref. 1.

### X-ray diffraction measurement

Figure S1 shows a XRD pattern of a CaK1144 sample presented on a log scale. All the peaks can be indexed by 00$l$ for CaK1144. The small peaks indicated by blue triangles arise from the residual Cu K$\beta$ radiation. No trace of CaFe$_2$As$_2$ (Ca122) and KFe$_2$As$_2$ (K122) phases was observed within the resolution of in-lab XRD.

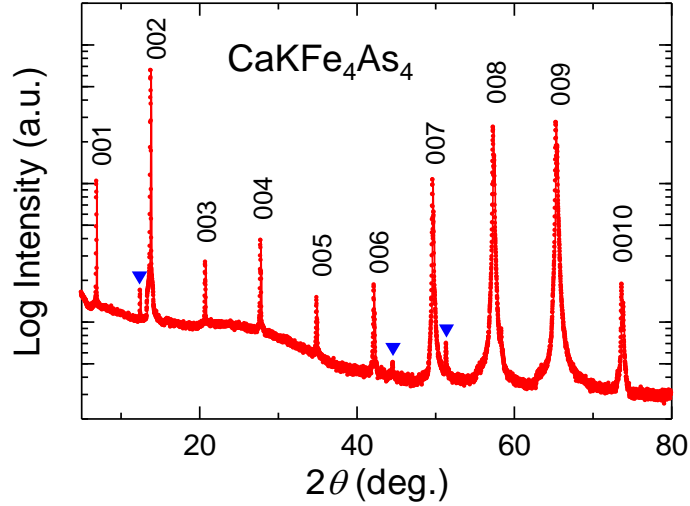

**Fig. S1** XRD patterns for a CaK1144 single crystal. Blue triangles indicate residual Cu K$\beta$ reflections.

### Magnetization and in-plane resistivity

Next, Figure S2a shows temperature ($T$) dependence of $M$ of CaK1144. $M$ was measured with zero-field-cooled and field-cooled processes under a magnetic field ($H$) of 1 mT along the $c$ axis. A sharp superconducting transition was observed around onset $T_c$ = 35.5 K. There is no trace of additional transition around $T$ = 3.8 K arising from K122. Figure S2b shows $T$ dependence of $\rho_{ab}$. $\rho_{ab}$ shows a metallic behavior characterized by a S-shaped curve, which is similar to that observed for optimally-doped BaK122 (see Figure S2f). No anomaly corresponding to magneto-structural transition of Ca122 was detected in $\rho_{ab}$. The residual resistivity ratio $\rho_{ab}$(300K)/$\rho_{ab}$(40K) is about 16, which is comparable with that for the best sample shown in Ref. 1. Thus, the results of XRD, $M$, and $\rho_{ab}$ measurements suggest that a "phase-pure" CaK1144 single crystal was obtained. We performed STEM experiments on such a selected crystal.

We also evaluated the anisotropy of upper critical field ($H_{c2}$). Figure S2c shows $\rho_{ab}$ of CaK1144 under $H$ along $c$ axis (upper panel) and $ab$ plane (lower panel) measured up to $\mu_0 H$ = 9 T. The superconducting transition shifts towered lower temperatures with increasing $H$. The broadening of the transition is weak, which can be associated with the small anisotropy of CaK1144. Figure S2d shows the $T$ dependence of $H_{c2}$ for $H$ // $c$ (blue) and $ab$ (green) obtained from 90% (filled) and 10% (open) criteria shown in Figure S2c. The slope (|d$H_{c2}$/d$T$|) was estimated

by a linear fit to 1-9 T data, and |d$H_{c2}$/d$T$| ~ 5.5 T/K and ~ 9.9 T/K were obtained for $H$ // $c$ and $ab$, respectively. Using the Werthamer-Helfand-Hohenberg formula ($H_{c2}(0) = 0.69T_c$|d$H_{c2}$/d$T$|), $H_{c2}(0)$ was estimated to be 140 T and 240 T for $H$ // $c$ and $ab$, respectively. The anisotropy factor ($\gamma$) is about 1.8. These superconducting properties are comparable with those for BaK122 (for comparison, the same data set as in Figures S2a-d is shown in Figures S2e-h).

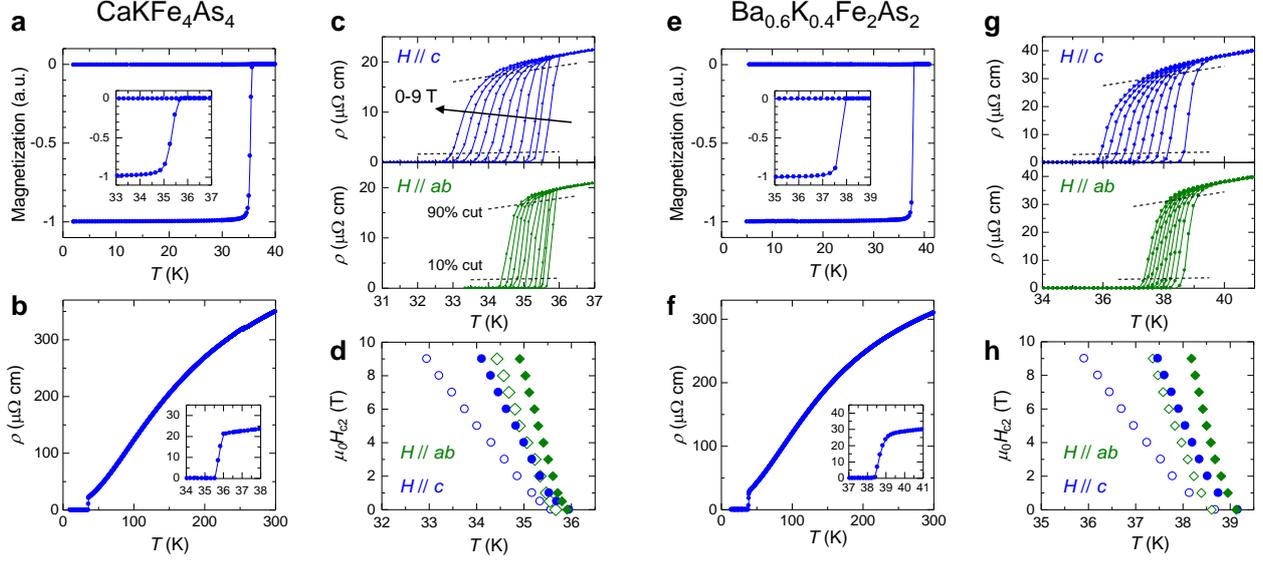

**Fig. S2** Characterization of CaK1144 and BaK122 single crystals. **a** $T$ dependence of $M$ for CaK1144. Inset shows the magnified view around the superconducting transition. **b** $T$ dependence of $\rho_{ab}$. Inset shows the magnified view around the superconducting transition. **c** $\rho_{ab}$ under the magnetic fields up to 9 T for $H$ // $c$ (upper panel) and $ab$ (lower panel). **d** $T$ dependence of $H_{c2}$ for $H$ // $c$ (blue) and $ab$ (green). Filled and open symbols indicate $H_{c2}$ determined by 90% and 10% criteria (shown in Figure S2c), respectively. **e-f** The same data set as shown in **a-d** for BaK122.

## 2. Chemical composition analysis of CaK1144

The chemical compositions of CaK1144 and defects were investigated by electron energy loss spectroscopy (EELS) and energy dispersive X-ray spectroscopy (EDS). Figure S3a shows the ADF-STEM image and EELS spectra around a defect in CaK1144 matrix (indicated by a black arrow). The alternation of Ca and K layers is violated; two Ca layers appears (or a K layer is skipped) around the defect. Meanwhile, no significant change was found for Fe and As. Similar results were obtained by EDS as shown in Figure S3b. In addition, the average chemical composition away from the defect is Ca : K : Fe : As = 11.0 : 10.7 : 41.2 : 37.1, which agrees with CaK1144 within the measurement error range of EDS.

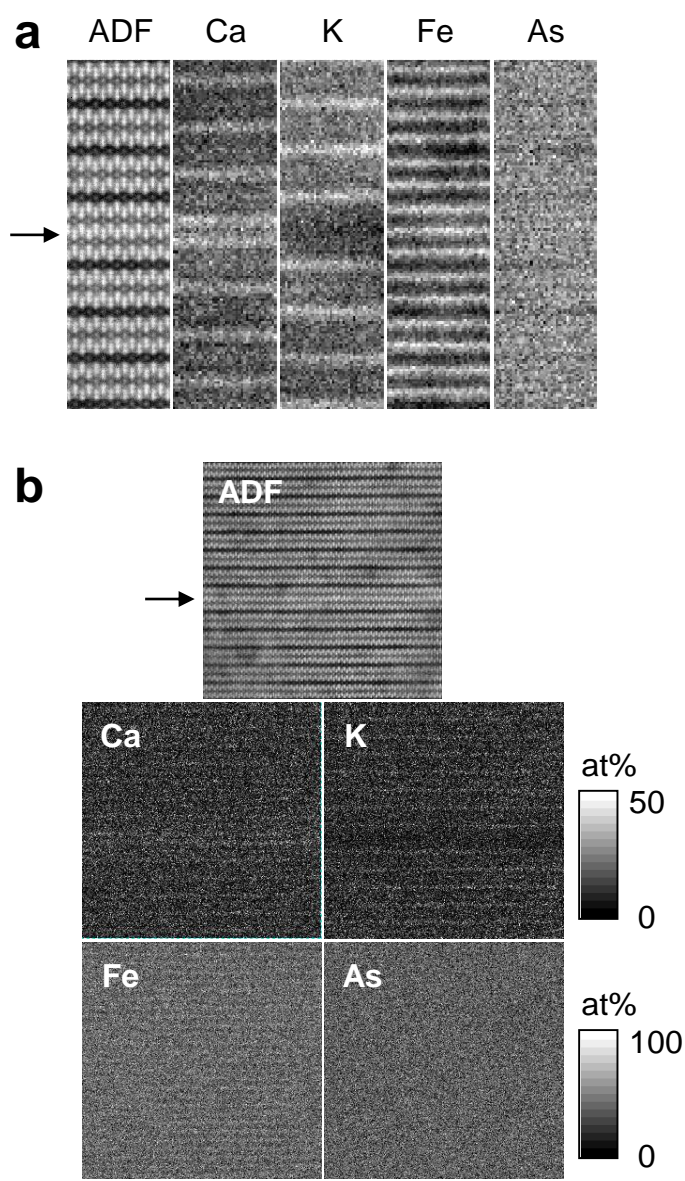

**Fig. S3** Chemical composition analysis of CaK1144. **a** ADF-STEM image and EELS spectra, and **b** ADF-STEM image and EDS mapping around a defect (indicated by black arrows) in CaK1144 matrix.

## 3. Temperature dependence of $J_c$ for $H \parallel c$

Figure S4a shows the $T$ dependence of $J_c$ for $H \parallel c$ ($J_c^{H//c} - T$) under 1, 1.4, and 2 T. The peak in $J_c^{H//c} - T$ ($T_p$) can be defined for $H > 1$ T, while it is absent at low $H$ below 1 T (a shoulder feature remains). This behavior can be understood by taking into account the two source of vortex pinning; the strong vortex pinning and the superconducting (SC) defects. Figure S4b shows the $J_c^{H//c} - T$ curves plotted in the double logarithmic scale. At low $T$ and $H$, $J_c^{H//c}$ shows a power-law decay ($J_c \sim H^{-0.7}$), which is observed in various iron-based superconductors and often associated with the strong vortex pinning arising from the sparse and large pointlike defects.[2] At low $H$, the strong pinning contribution, which monotonically decreases with $T$, is likely dominant, resulting in the monotonous decrease of $J_c$. On the other hand, since the strong pinning contribution rapidly decreases in $H$, the pinning arising from the SC defects (thin Ca122 intergrowths) which is enhanced with increasing $T$ becomes dominant at higher $H$, thus $T_p$ can be clearly observed.

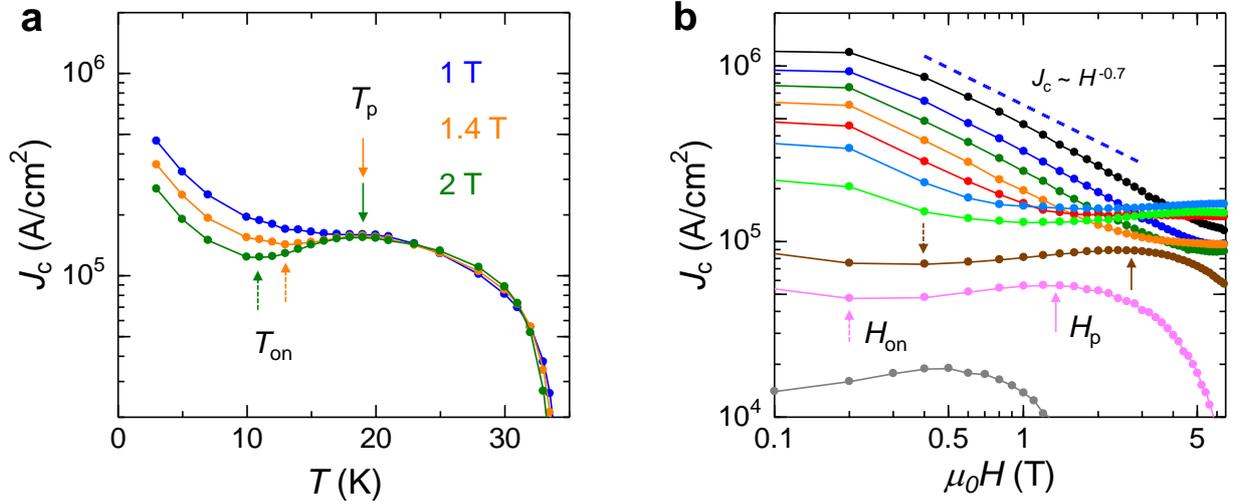

**Fig. S4** $T$ and $H$ dependence of $J_c^{H//c}$. **a** $T$ dependence of $J_c^{H//c}$ under 1, 1.4, and 2 T. The peak ($T_p$) and local minimum ($T_{on}$) are indicated by solid and dashed arrows, respectively. **b** $H$ dependence of $J_c^{H//c}$ plotted in the double logarithmic scale. The dashed line shows a power-law relation, $J_c \sim H^{-0.7}$. The peak ($H_p$) and local minimum ($H_{on}$) are indicated by solid and dashed arrows, respectively.

## 4. Calculation of pinning force density

The pinning force density ($f_p$) was calculated using a simple model. For pinning centres larger than coherence length ($\xi$), the pinning energy ($E_p$) is proportional to $\Delta E_c \xi^2$ where $\Delta E_c$ is the difference in condensation energy between SC defect and Ca1144 matrix, hence we obtain $f_p \sim E_p/\xi \sim \Delta E_c \xi$. Here, the $T$ dependences of $E_c$ and $\xi$ are modelled by $E_c \sim H_c^2 \sim [1 - (T/T_c)^2]^2$ ($H_c$: thermodynamic critical field) and $\xi \sim [(1 + (T/T_c)^2)/(1 - (T/T_c)^2)]^{1/2}$, respectively. In Figure S5, the calculation results using different $T_c^{\text{defect}}$ and $E_c^{\text{defect}}(0)/E_c^{1144}(0)$ values are shown. It can be seen that the peak position of $f_p$ shifts to lower $T$ when $T_c^{\text{defect}}$ is lowered and that the peak is more prominent when $E_c^{\text{defect}}(0)/E_c^{1144}(0)$ is closer to 1. This simple model suggests that $T_c^{\text{defect}}$ is correlated with $T_p$ in $J_c^{H//c} - T$ curves.

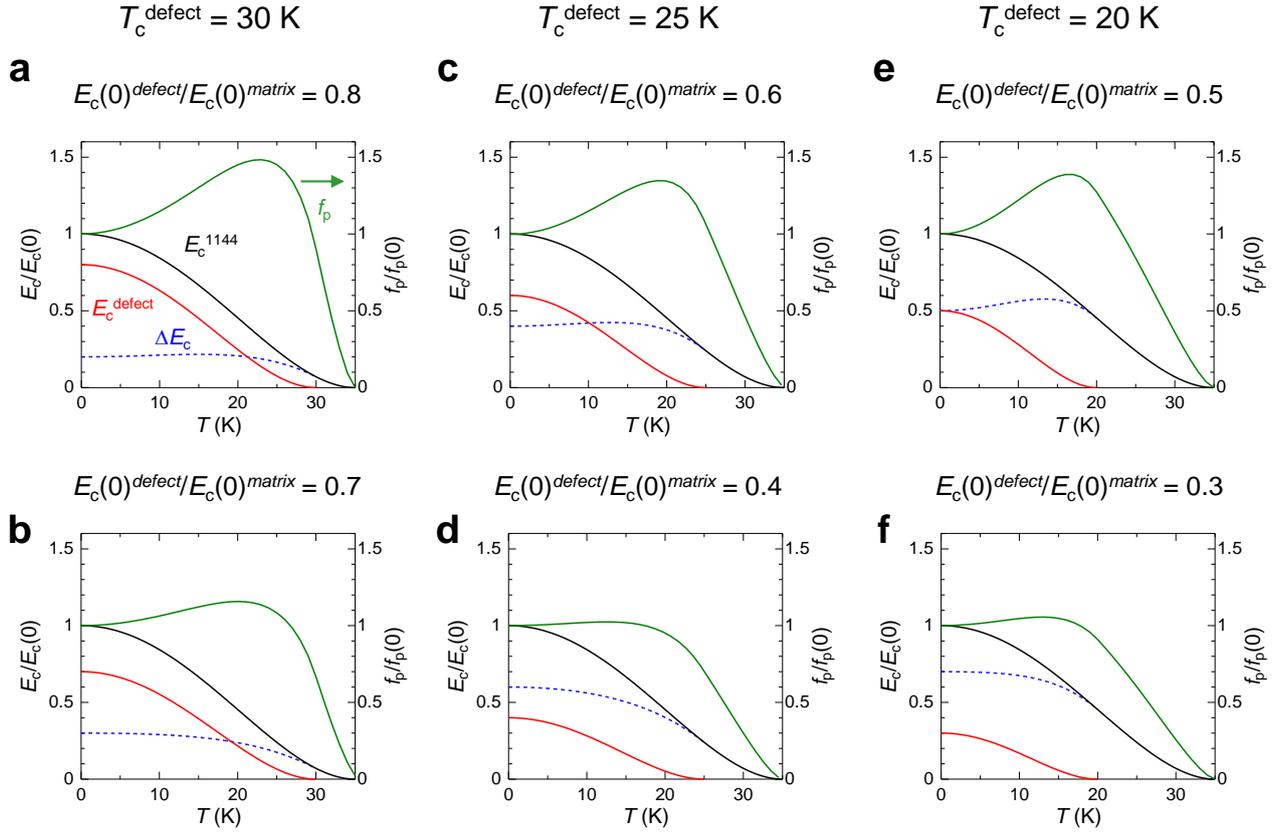

**Fig. S5** Temperature dependence of pinning force density $f_p$. **a,b** $f_p$ calculated using $T_c^{\text{defect}}$ = 30 K and $E_c^{\text{defect}}(0)/E_c^{1144}(0)$ = 0.8 and 0.7, **c,d** $T_c^{\text{defect}}$ = 25 K and $E_c^{\text{defect}}(0)/E_c^{1144}(0)$ = 0.6 and 0.4, and **e,f** $T_c^{\text{defect}}$ = 20 K and $E_c^{\text{defect}}(0)/E_c^{1144}(0)$ = 0.5 and 0.3, respectively.

## 5. Calculation of anisotropic $J_c$ for $H // ab$

In the manuscript, we evaluated $J_c^{H//ab}$ using a simplified formula by taking $J_c^{H//ab} = J_c^c$. Meanwhile, to evaluate $J_c^{H//ab}$ more properly, two $J_c$ components, i.e. $J_c^{H//ab}$ and $J_c^c$ should be considered. Here, we describe the calculation procedure of $J_c^{H//ab}$ and $J_c^c$ based on the extended Bean's critical state model for anisotropic $J_c$.[3,4]

First, because there are two $J_c$ components, we need to perform two independent $M - H$ measurements on a rectangular sample with dimensions of $l$, $w$ ($< l$), and $d$ ($// c$) using different setups as shown in Figure S6.

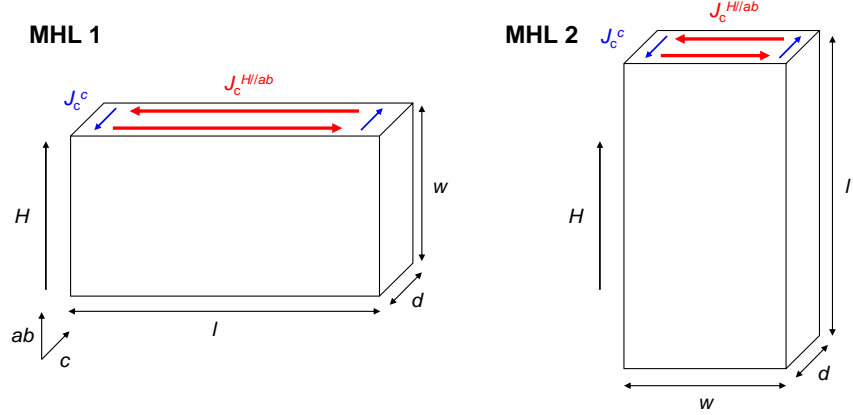

**Fig. S6** Two independent $M - H$ measurement setups for the evaluation of anisotropic $J_c$, $J_c^{H//ab}$ and $J_c^c$, using extended Bean's critical state model. $H$ is applied along $w$ (MHL 1) and $l$ (MHL 2).

Next, as described in the reference paper,[3] different current-flow configurations should be adopted for the calculation depending on the relation between the sample shape ($l/d$ or $w/d$) and the magnitude of $J_c$ anisotropy ($J_c^{H//ab}/J_c^c$). The possible configurations are shown in Figure S7 together with the conditions of $J_c^{H//ab}/J_c^c$ for each pattern.

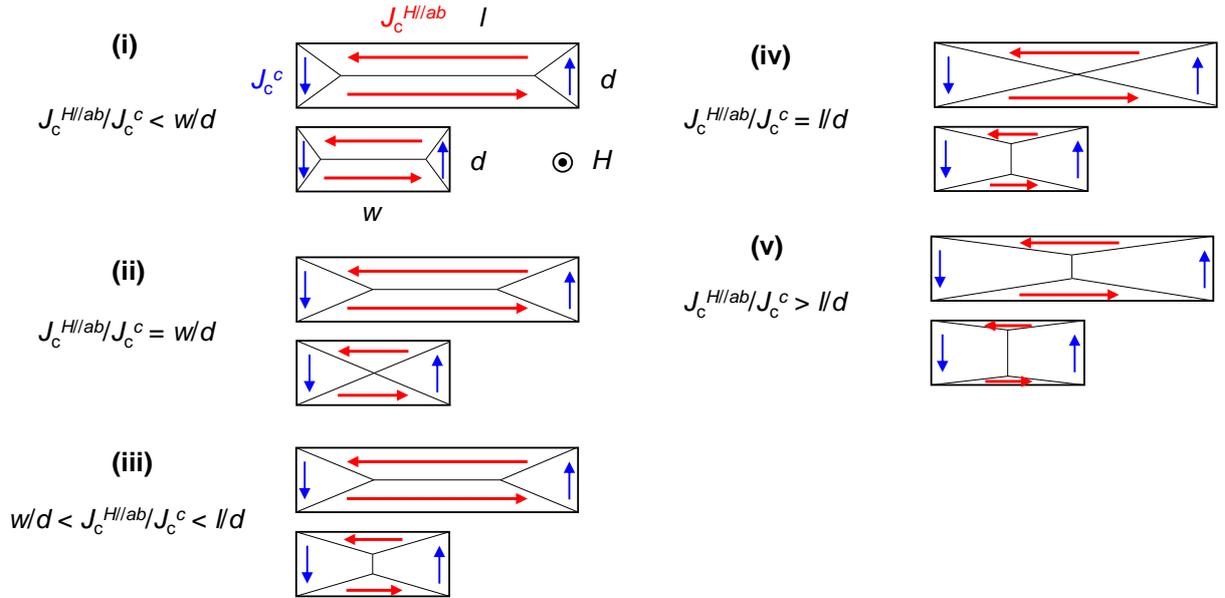

**Fig. S7** Possible configurations of the current flows for $H // ab$. The conditions of $J_c^{H//ab}/J_c^c$ for each configuration are also shown.

Based on these configurations, the width of MHL, $\Delta M_1$ and $\Delta M_2$ for MHL 1 and MHL 2, respectively, can be expressed by the following equations:

$$\Delta M_1 = \begin{cases} \dfrac{d\, J_c^{H//ab}}{20}\left(1 - \dfrac{d\, J_c^{H//ab}}{3l\, J_c^c}\right) & \text{(for (i) and (iii))} \\[1em] \dfrac{l\, J_c^c}{20}\left(1 - \dfrac{l\, J_c^c}{3d\, J_c^{H//ab}}\right) & \text{(for (v))} \end{cases}$$

$$\Delta M_2 = \begin{cases} \dfrac{d\, J_c^{H//ab}}{20}\left(1 - \dfrac{d\, J_c^{H//ab}}{3w\, J_c^c}\right) & \text{(for (i))} \\[1em] \dfrac{w\, J_c^c}{20}\left(1 - \dfrac{w\, J_c^c}{3d\, J_c^{H//ab}}\right) & \text{(for (iii) and (v))} \end{cases}$$

Using $\Delta M_1$ and $\Delta M_2$, $J_c^{H//ab}$ and $J_c^c$ for cases (i), (iii), and (v) in Figure S7 can be calculated as follows (here we define $\alpha = w/l\ (<1)$):

For (i): $J_c^{H//ab}/J_c^c < w/d$

$$J_c^{H//ab} = \frac{20\,(\Delta M_1 - \alpha \Delta M_2)}{t\,(1-\alpha)}$$

$$J_c^c = \frac{20\,(\Delta M_1 - \alpha \Delta M_2)^2}{3w\,(1-\alpha)(\Delta M_1 - \Delta M_2)}$$

For (iii): $w/d < J_c^{H//ab}/J_c^c < l/d$

$$\left[\frac{\alpha d^3}{3} - \frac{\alpha^2 d^3}{27}\right] J_c^{H//ab\,3} - 20\left[\frac{\alpha d^2}{3}\Delta M_1 + d^2 \Delta M_2\right] J_c^{H//ab\,2}$$
$$+ 800 d\, \Delta M_1 \Delta M_2\, J_c^{H//ab} - 8000\Delta M_1^2 \Delta M_2 = 0$$

$$\left[\frac{w^3}{3} - \frac{\alpha w^3}{27}\right] J_c^{c\,3} - 20\left[w^2 \Delta M_1 + \frac{w^2}{3}\Delta M_2\right] J_c^{c\,2}$$
$$+ 800 w\, \Delta M_1 \Delta M_2\, J_c^c - 8000\Delta M_1 \Delta M_2^2 = 0$$

For (v): $l/d < J_c^{H//ab}/J_c^c$

$$J_c^{H//ab} = \frac{20\,(\Delta M_2 - \alpha^2 \Delta M_1)^2}{3\alpha d\,(1-\alpha)(\Delta M_2 - \alpha \Delta M_1)}$$

$$J_c^c = \frac{20\,(\Delta M_2 - \alpha^2 \Delta M_1)}{w\,(1-\alpha)}$$

Then, in order to calculate $J_c^{H//ab}$ and $J_c^c$ from the experimental data, the equations (cases (i), (iii), and (v)) should be appropriately chosen. To determine which equations to use, $\Delta M_1/\Delta M_2$ is a useful parameter which can be easily obtained from the data. From the expressions of $\Delta M_1$ and $\Delta M_2$ shown before, $\Delta M_1/\Delta M_2$ for each case can be described as follows:

$$\frac{\Delta M_1}{\Delta M_2} = \begin{cases} \dfrac{1 - \dfrac{d\, J_c^{H//ab}}{3l\, J_c^c}}{1 - \dfrac{d\, J_c^{H//ab}}{3w\, J_c^c}} & (J_c^{H//ab}/J_c^c < w/d \text{ ; for (i))} \\[2em] \dfrac{d\, J_c^{H//ab}}{w\, J_c^c} \dfrac{1 - \dfrac{d\, J_c^{H//ab}}{3l\, J_c^c}}{1 - \dfrac{w\, J_c^c}{3d\, J_c^{H//ab}}} & (w/d < J_c^{H//ab}/J_c^c < l/d \text{ ; for (iii))} \\[2em] \dfrac{l}{w} \dfrac{1 - \dfrac{l\, J_c^c}{3d\, J_c^{H//ab}}}{1 - \dfrac{w\, J_c^c}{3d\, J_c^{H//ab}}} & (l/d < J_c^{H//ab}/J_c^c \text{ ; for (v))} \end{cases}$$

For the specific cases, (ii) and (iv), $\Delta M_1/\Delta M_2$ becomes

$$\frac{\Delta M_1}{\Delta M_2} = \begin{cases} \dfrac{3 - \alpha}{2} & (J_c^{H//ab}/J_c^c = w/d \text{ ; case (ii))} \\[1em] \dfrac{2}{\alpha(3 - \alpha)} & (J_c^{H//ab}/J_c^c = l/d \text{ ; case (iv))} \end{cases}$$

Thus, one can choose cases (i), (iii), and (v) for $\Delta M_1/\Delta M_2 < (3-\alpha)/2$, $(3-\alpha)/2 < \Delta M_1/\Delta M_2 < 2/\alpha(3-\alpha)$, and $2/\alpha(3-\alpha) < \Delta M_1/\Delta M_2$, respectively.

Below, we show the calculation results for a CaK1144 sample with dimensions of $l = 1.26$ mm, $w = 0.80$ mm, and $d = 0.064$ mm. Figure S8a shows the two MHLs, $M_1$ and $M_2$, obtained from different setups. Clearly, $\Delta M_1$ and $\Delta M_2$ are different from each other ($\Delta M_1 > \Delta M_2$). Here, the values of $(3-\alpha)/2$ and $2/\alpha(3-\alpha)$ are 1.18 and 1.33 ($\alpha = w/l = 0.635$), respectively, and $\Delta M_1/\Delta M_2$ takes 1.19-1.26, thus the case (iii) (($3-\alpha)/2 < \Delta M_1/\Delta M_2 < 2/\alpha(3-\alpha)$)) was chosen. Figure S8b shows the obtained $J_c^{H//ab}$ (red circles) and $J_c^c$ (blue circles) using the equations for case (iii). The anisotropy of $J_c$, $J_c^{H//ab}/J_c^c$, is found to be 13-15, confirming a large anisotropy. Thus, we successfully derived $J_c^{H//ab}$ and $J_c^c$ based on the extended Bean's critical state model.

In addition, we calculated $J_c^{H//ab}$ from each single MHL ($M_1$ or $M_2$) using the simplified formula by taking $J_c^{H//ab} = J_c^c$. The $J_c^{H//ab}$ values from $M_1$ and $M_2$ are shown in Figure S8b (open circles and triangles, respectively). Owing to the averaging of two components, $J_c^{H//ab}$ is underestimated. Meanwhile, it can be seen that $J_c^{H//ab}$ derived from $M_1$ (open circles) is closer to $J_c^{H//ab}$ from the extended Bean model (red circles) compared with $J_c^{H//ab}$ derived from $M_2$ (open triangles), indicating that the effect of anisotropy on evaluation of $J_c^{H//ab}$ becomes less significant when $l \gg d$. The data shown in the manuscript were obtained for the sample with $l/d = 45$, which is larger than that of the present sample ($l/d = 20$), hence the contribution from $J_c^c$ should be practically negligible. Thus, we confirmed that $J_c^{H//ab}$ was reasonably evaluated by the simplified calculation and that the procedure does not alter the discussion nor conclusions in this study.

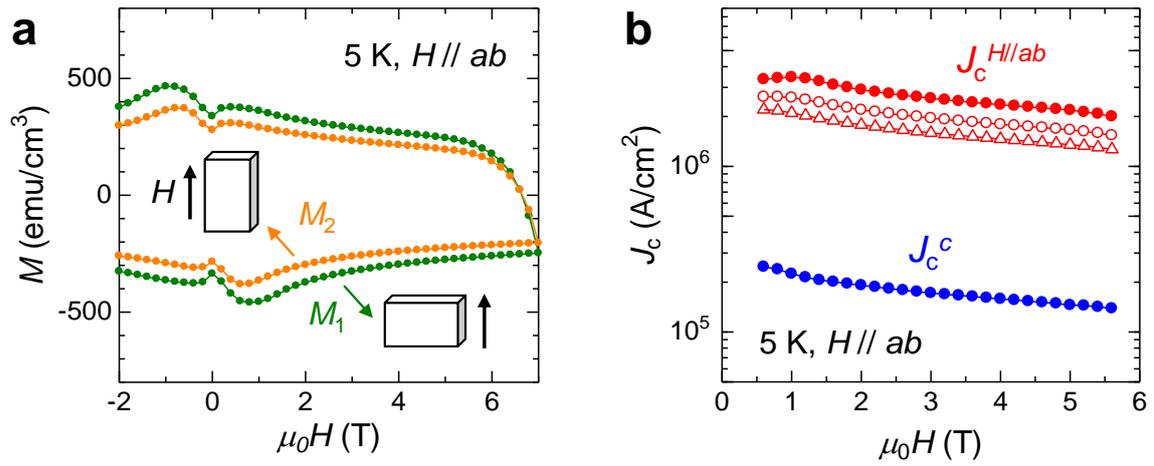

**Fig. S8** Evaluation of $J_c^{H//ab}$ (red circles) and $J_c^c$ for CaK1144. **a** MHLs for two setups ($M_1$ (green, $H // w$) and $M_2$ (orange, $H // l$)) at 5 K under $H // ab$. **b** $H$ dependence of $J_c^{H//ab}$ (red circles) and $J_c^c$ (blue circles) derived from $M_1$ and $M_2$ using the extended Bean's critical state model.[3,4] The open circles and triangles are $J_c^{H//ab}$ values calculated using the simplified formula by taking $J_c^{H//ab} = J_c^c$ from each single MHL, $M_1$ and $M_2$, respectively.